\documentclass[aps,twocolumn,prb,10pt]{revtex4-2}
\usepackage{graphicx}
\usepackage{amsmath}
\usepackage{amssymb}
\usepackage{bm}
\usepackage{color}

\def\l{\langle}
\def\r{\rangle}
\def\Tr{{\rm Tr}}
\def\kB{{k_{\rm B}}}
\newcommand{\CHK}[1]{{\textcolor{black}{#1}}}

\def\myfigOne{
\begin{figure}
\begin{center}
\includegraphics[width=8cm]{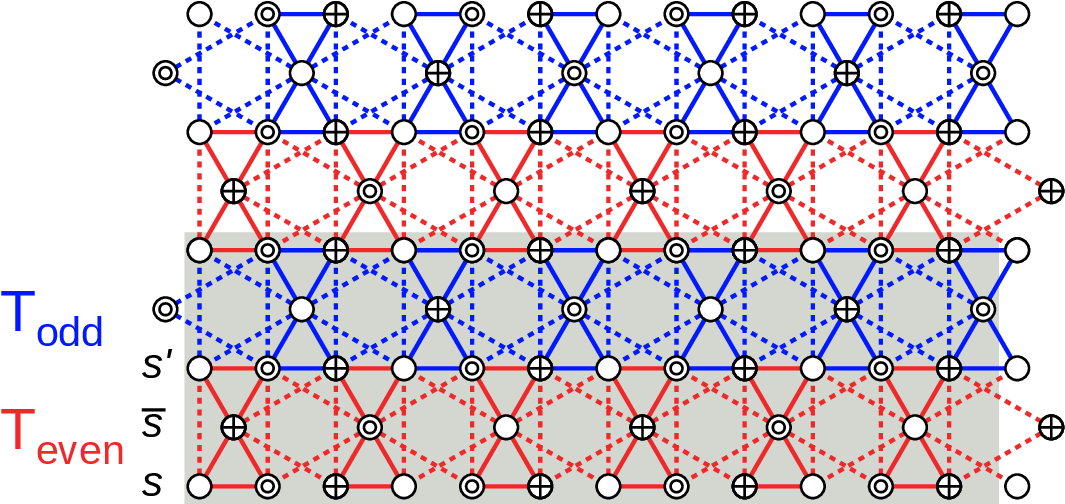}
\caption{
 The schematic representation of Hamiltonian~(\ref{eq_H}) on the kagome lattice.
 Open, double, and cross circles denote sites in the three sublattices, $\Lambda_0$, $\Lambda_1$, and $\Lambda_2$, respectively.
 The NN couplings $J_1$ are denoted by solid lines, and the NNN couplings $J_2$ are denoted by dashed lines. 
 The shaded area indicates the unit region of the $L=12$ system with a periodic boundary condition. 
 The row-to-row transfer matrix ${\bf T}_{\rm even}(s,s')$ [${\bf T}_{\rm odd}(s,s')$] accounts for the Boltzmann weights denoted by red (blue) lines; see the text.   
}
\label{fig_KagomeLattice}
\end{center}
\end{figure}
}
\def\myfigTwo{
\begin{figure}
\begin{center}
\includegraphics[width=8cm]{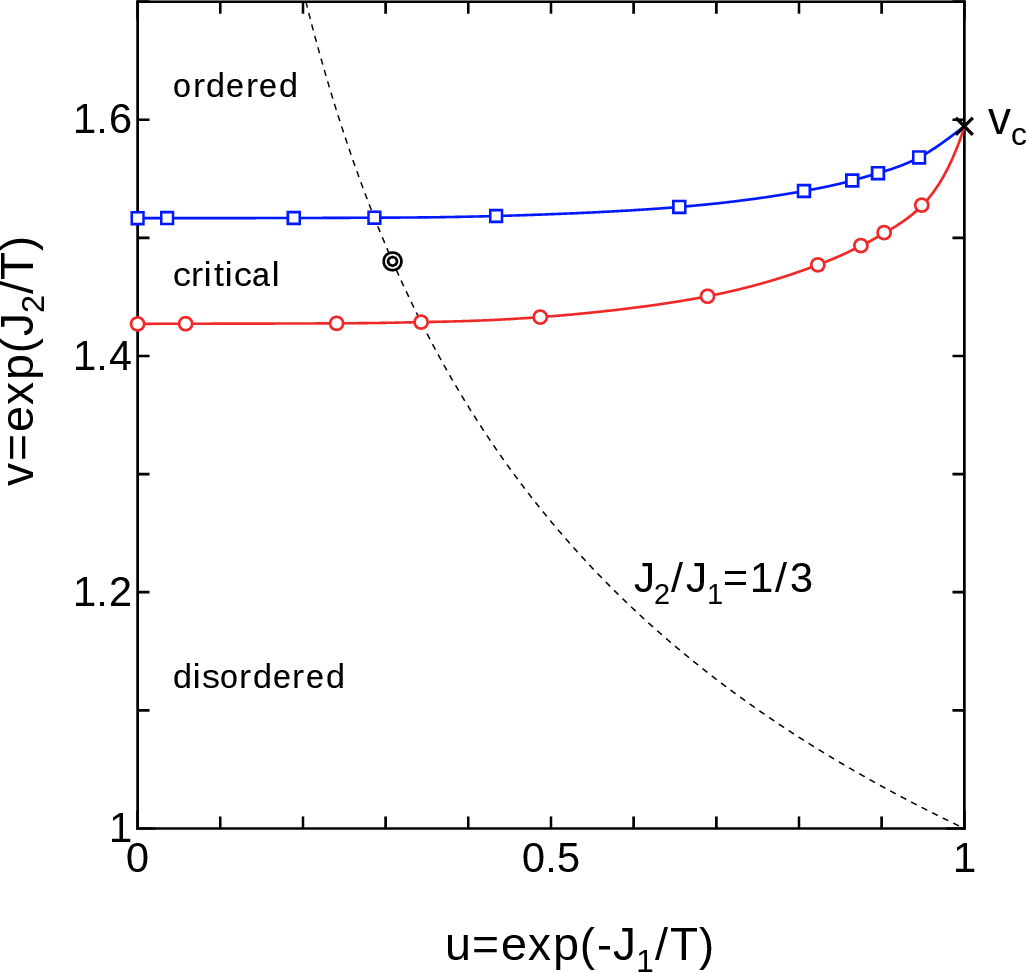}
\caption{
 The global phase diagram of the kagome-lattice AF Ising model with NNN F couplings on the $(u,v)$ plane \CHK{which consists of the ordered phase with sublattice magnetizations, the critical phase, and the disordered phase}.
 The blue squares (red circles) with a curve plot the BKT transition points $T_1$ ($T_2$) for the finite ratios $r=1/8$, 1/4, 1/3, 1/2, 1.0, 2.0, 3.0, 4.0, 8.0 and for the limiting case of $u=0$. 
 The cross mark, $\times$, on $v$-biaxis denotes the exact phase transition point of the kagome-lattice F Ising model $v_c=(3+2\sqrt3)^{\frac14}$.
 The dotted line satisfies the condition $r=1/3$; the double circle in the critical phase denotes the point $T=0.85$ on the line.
}
\label{fig_phase}
\end{center}
\end{figure}
}
\def\myfigThree{
\begin{figure}
\begin{center}
\includegraphics[width=8cm]{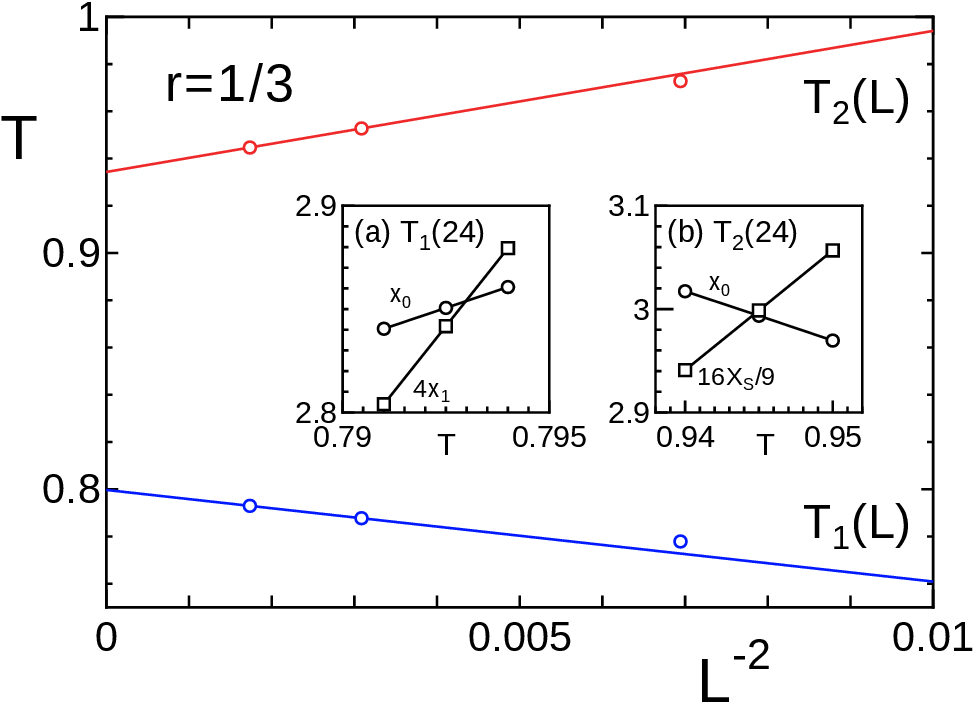}
\caption{
 The system-size dependence of $T_1(L)$ and $T_2(L)$ for $r=1/3$. 
 We draw the straight lines by fitting the $L=18$ and 24 data.  
 Insets (a) and (b) display, respectively, the level-cross conditions Eq.~(\ref{eq_LS1}) and Eq.~(\ref{eq_LS2}) for the scaled gaps of $L=24$ as functions of $T$.
}
\label{fig_extrap1}
\end{center}
\end{figure}
}
\def\myfigFour{
\begin{figure}
\begin{center}
\includegraphics[width=8cm]{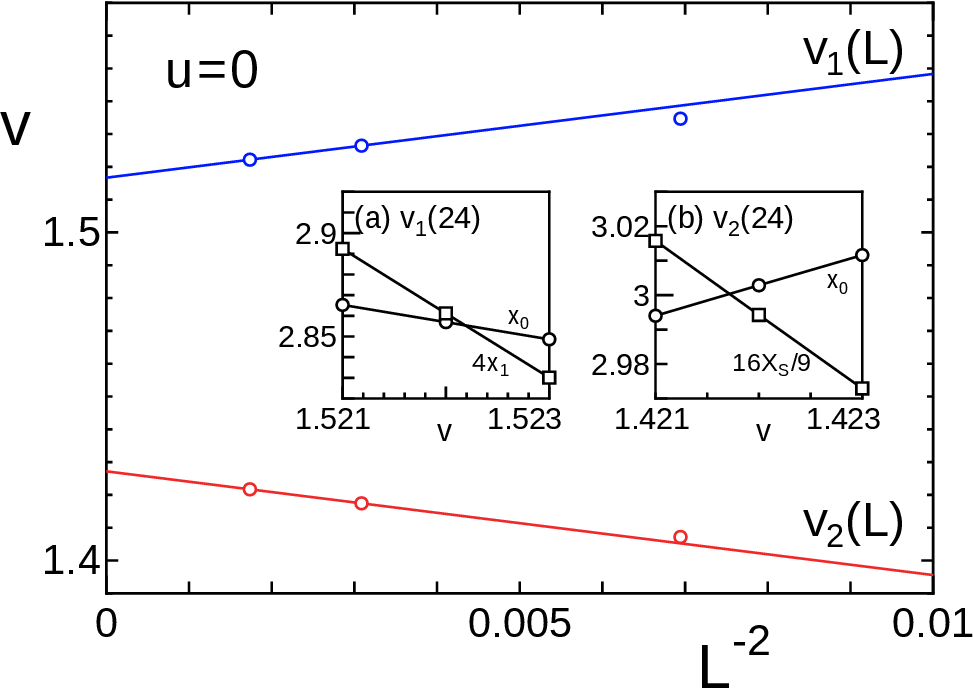}
\caption{
 The system-size dependence of $v_1(L)$ and $v_2(L)$ for $u=e^{-J_1/T}=0$. 
 We draw straight lines by fitting the $L=18$ and 24 data.  
 Insets (a) and (b) respectively display the level-crossing conditions Eq.~(\ref{eq_LS1}) and Eq.~(\ref{eq_LS2}) for the scaled gaps of $L=24$ as functions of $v=e^{+J_2/T}$.
}
\label{fig_extrap2}
\end{center}
\end{figure}
}
\def\myfigFive{
\begin{figure}
\begin{center}
\includegraphics[width=8cm]{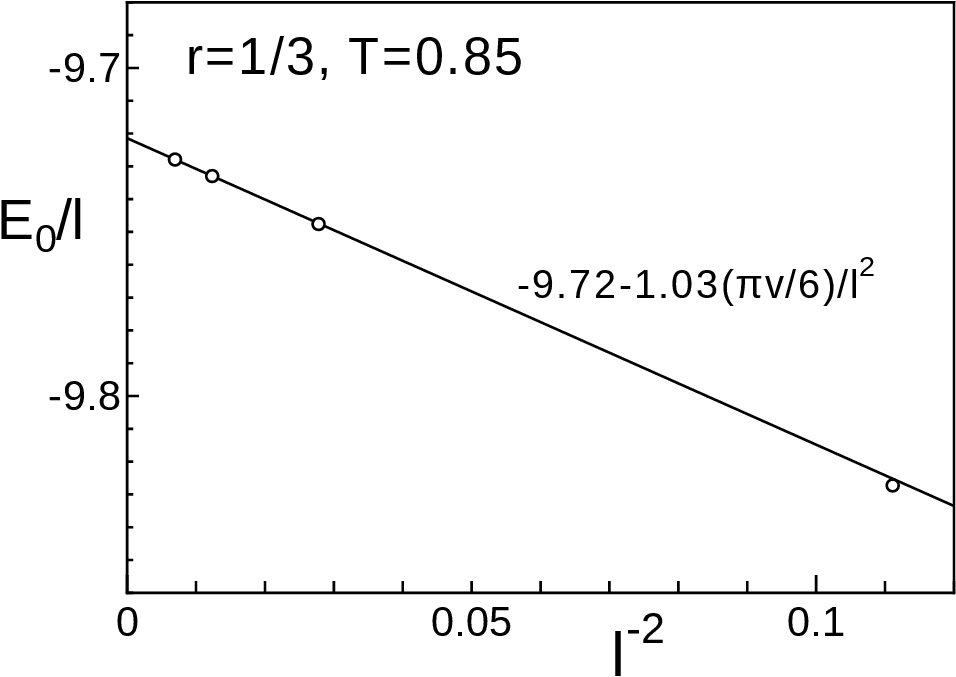}
\caption{
 The finite-size dependence of the largest eigenvalue of the TM for $r=1/3$ and $T=0.85$; see Fig.~\ref{fig_phase}.
 We fit the two largest size data, $l=L/2=9$ and 12, with an inverse geometric factor $v=\sqrt{3}$. 
 The slope of the fitting line estimates $c\simeq1.03$. 
}
\label{fig_c}
\end{center}
\end{figure}
}
\def\myfigSix{
\begin{figure}
\begin{center}
\includegraphics[width=8.1cm]{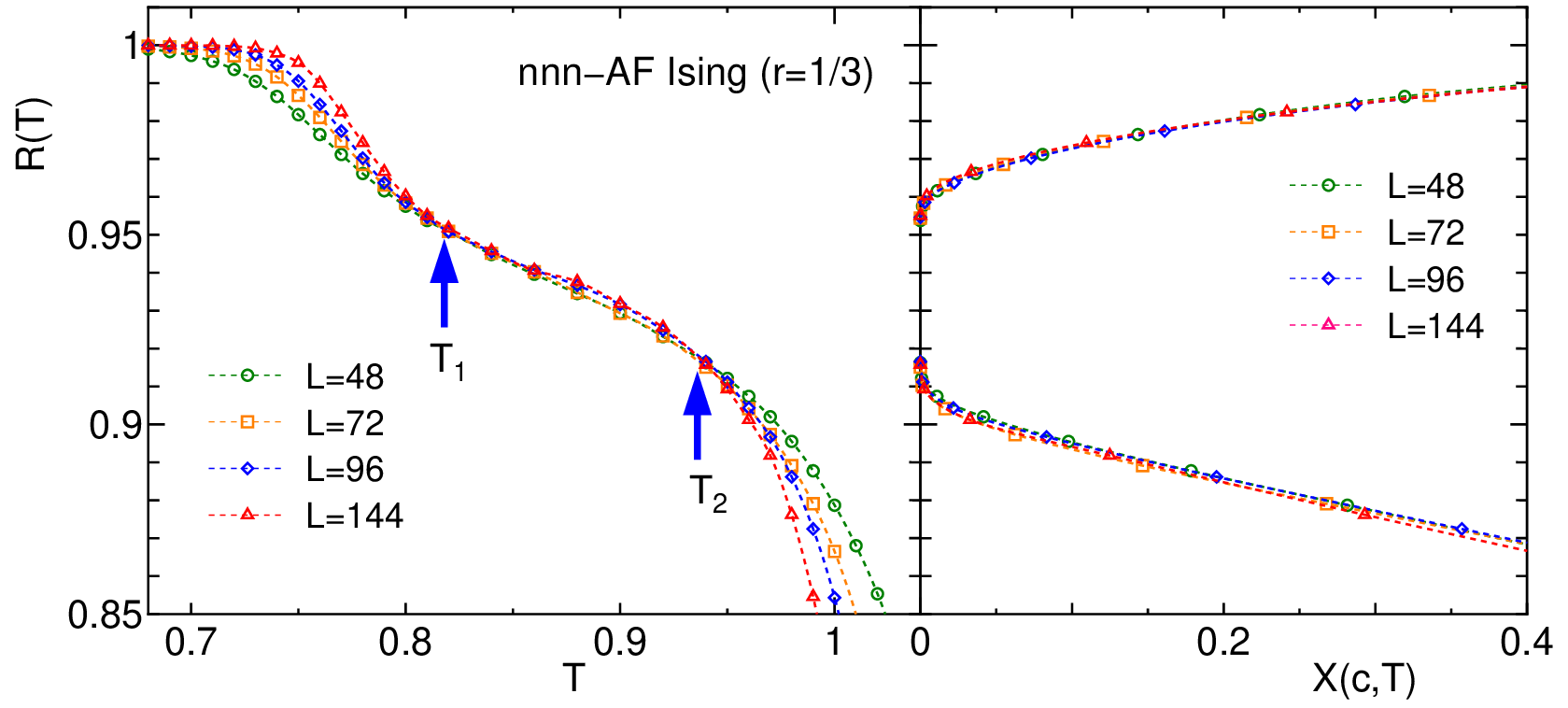}
\vspace*{2mm}

\includegraphics[width=8.1cm]{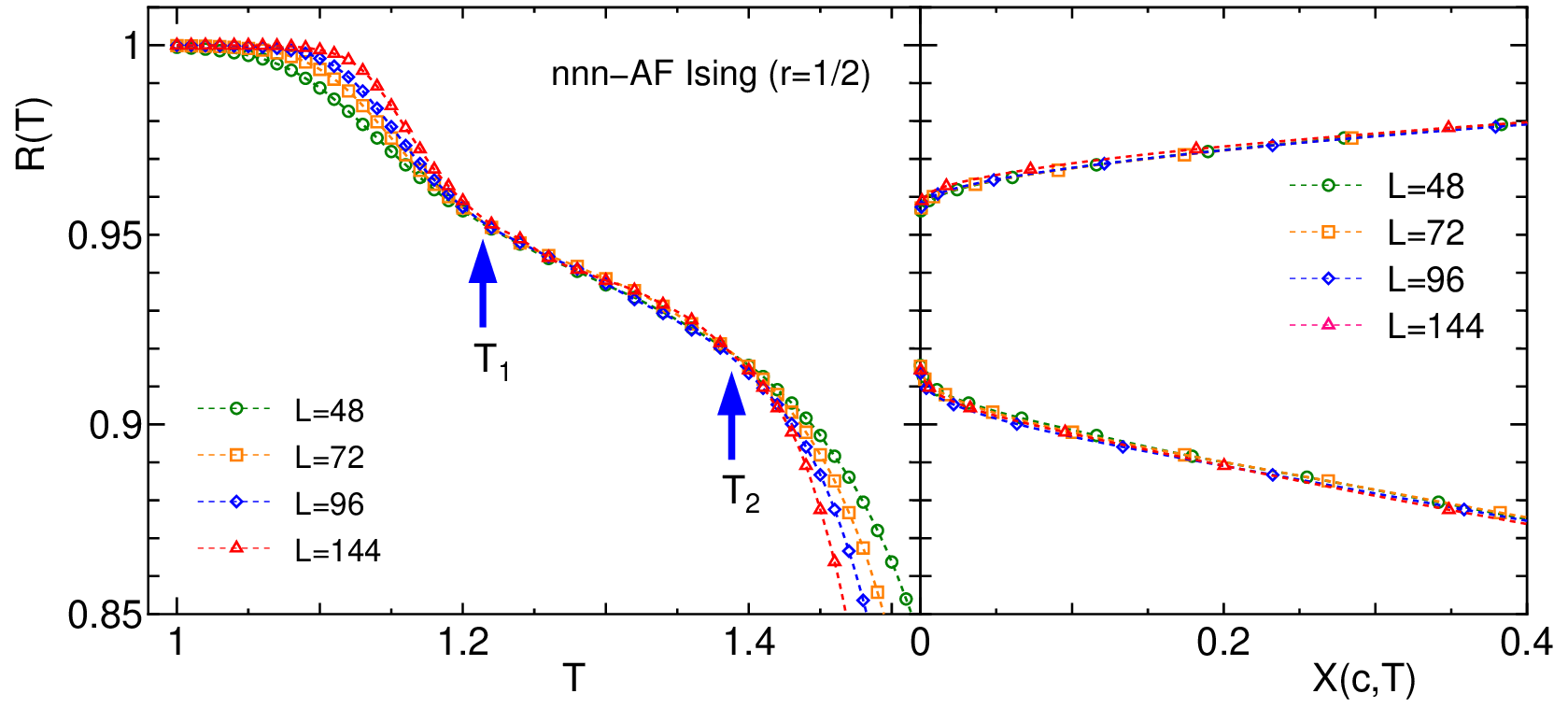}
\vspace*{2mm}

\includegraphics[width=8.1cm]{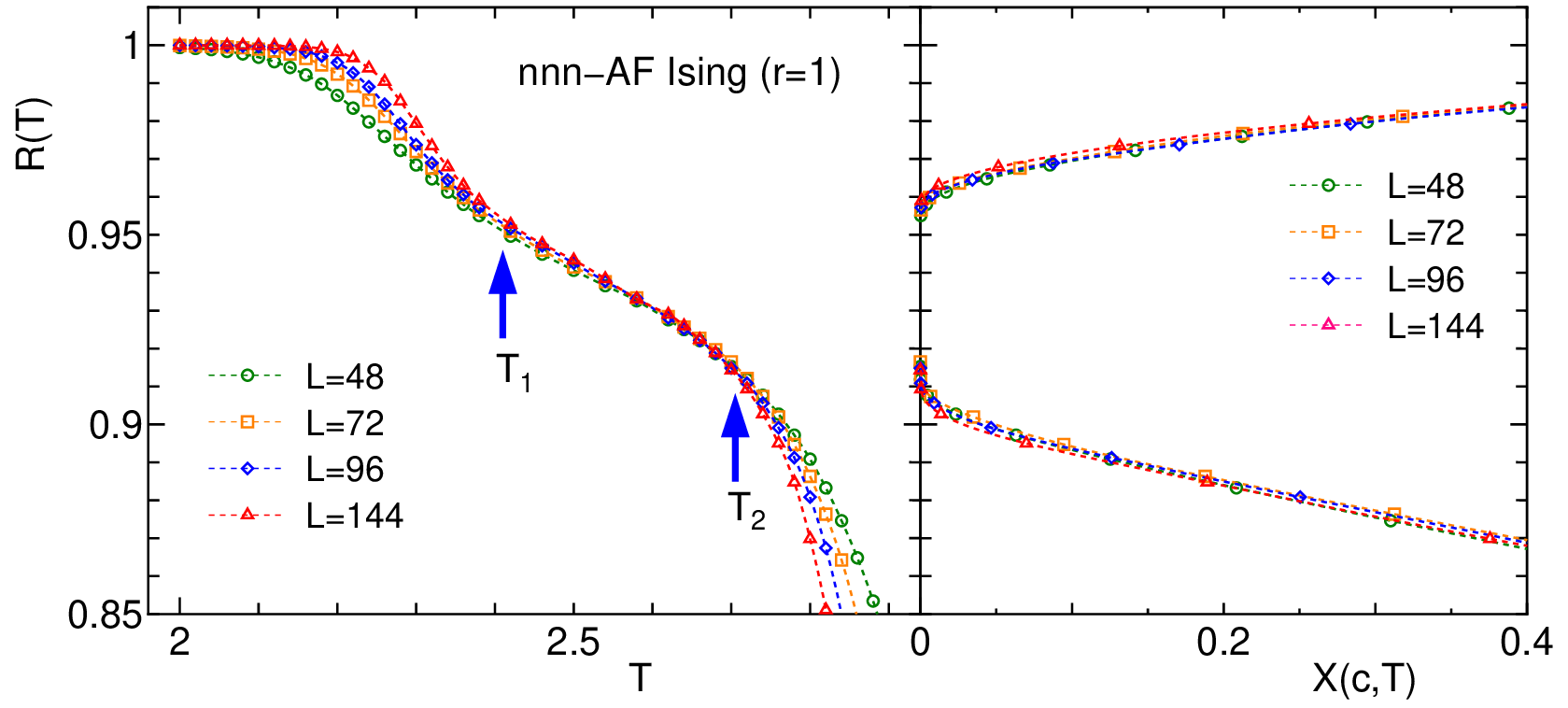}
\vspace*{2mm}

\includegraphics[width=8.1cm]{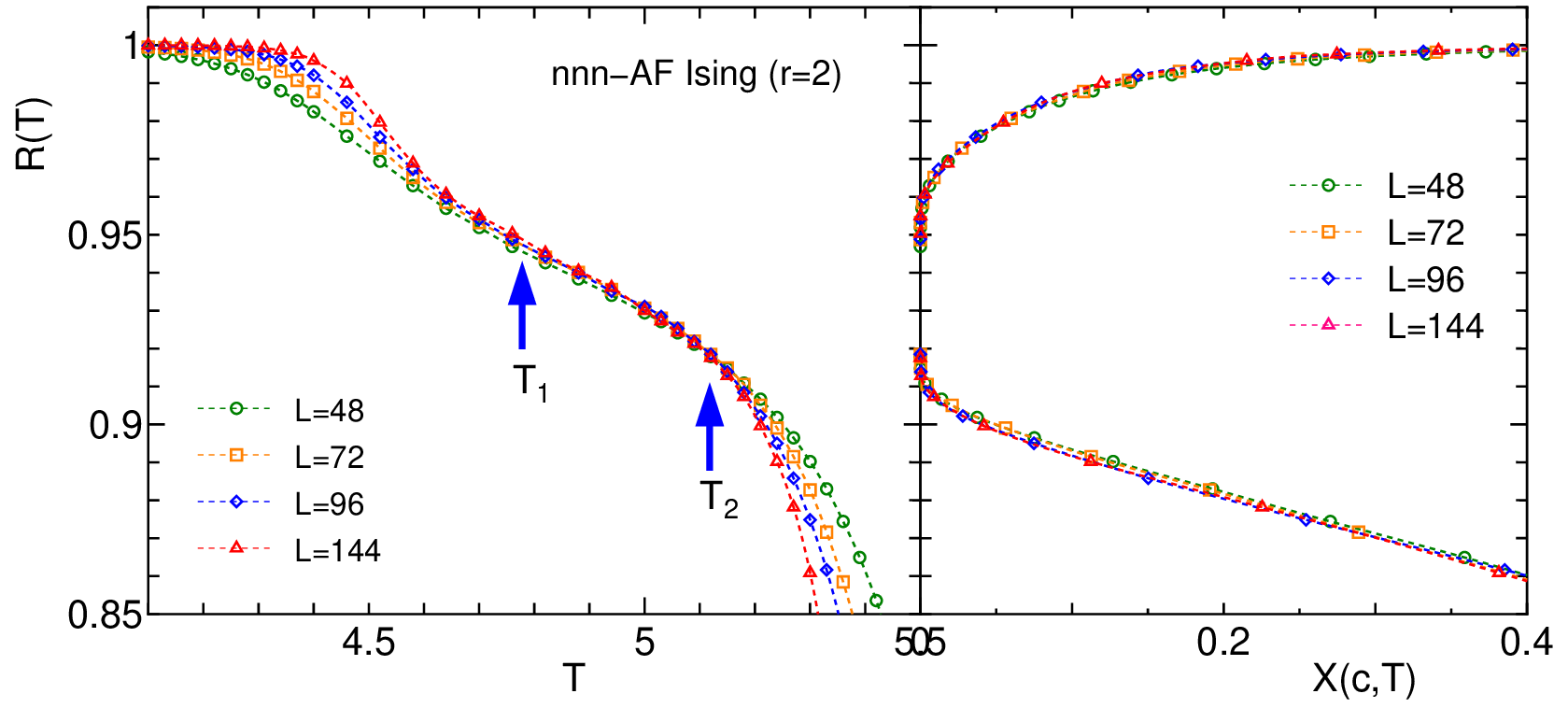}
\caption{
 The plot of the correlation ratio $R(T)$ for the AF Ising model with F NNN coupling ($r=1/3$, 1/2, 1.0, and 2.0) on the kagome lattice. 
 The system sizes are $L=48$, 72, 96, and 144.
 In the right panel, the FSS plots are given, where $X(c,T)=L/\exp(c_{1,2}/\sqrt{|T-T_{1,2}|})$. 
 For convenience, the locations of $T_1$ and $T_2$ are represented by arrows.
}
\label{fig_MC}
\end{center}
\end{figure}
}
\def\myfigSeven{
\begin{figure}
\begin{center}
\includegraphics[width=6.8cm]{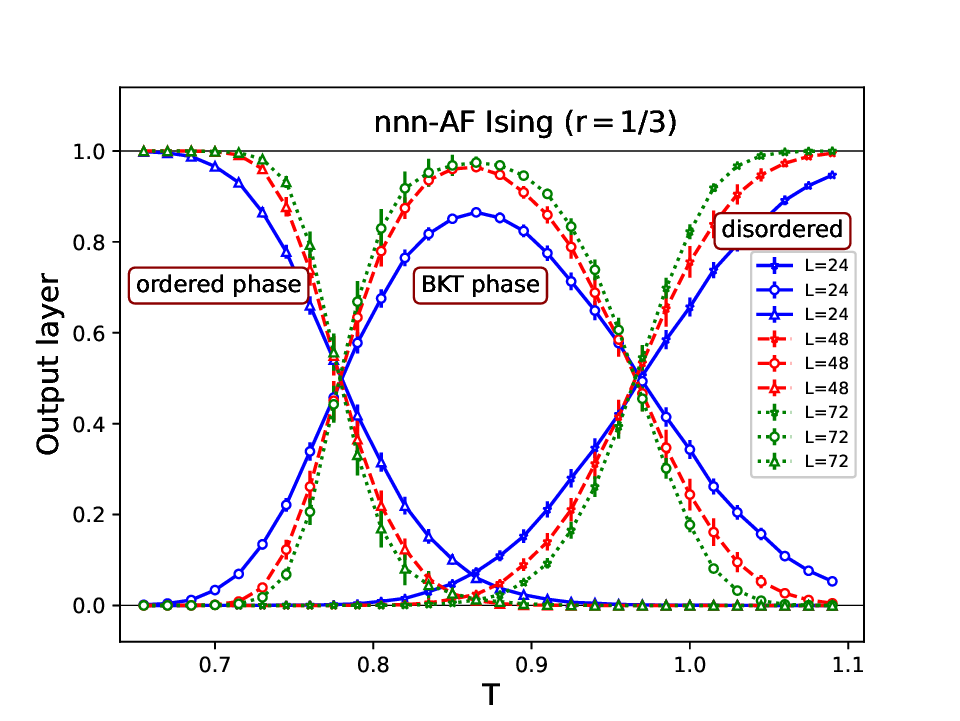}
\vspace*{2mm}

\includegraphics[width=6.8cm]{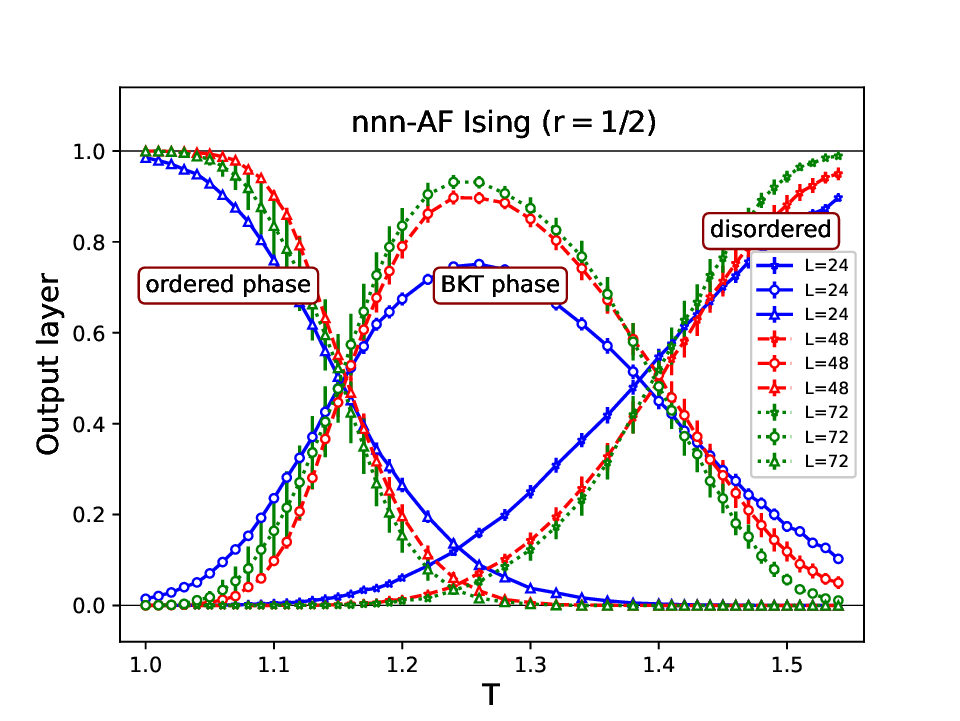}
\vspace*{2mm}

\includegraphics[width=6.8cm]{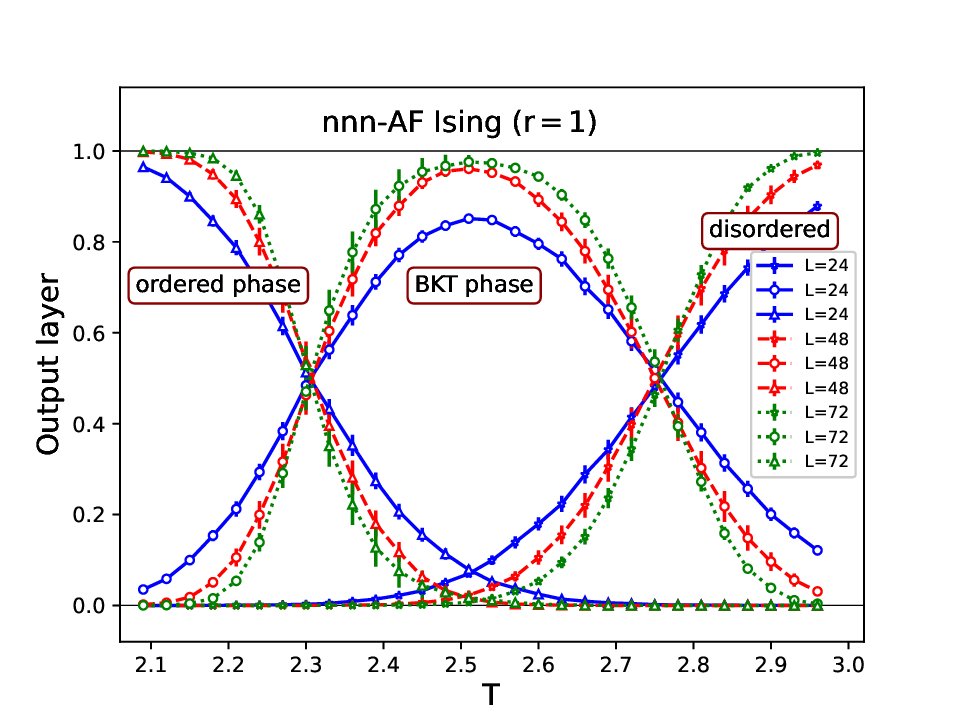}
\vspace*{2mm}

\includegraphics[width=6.8cm]{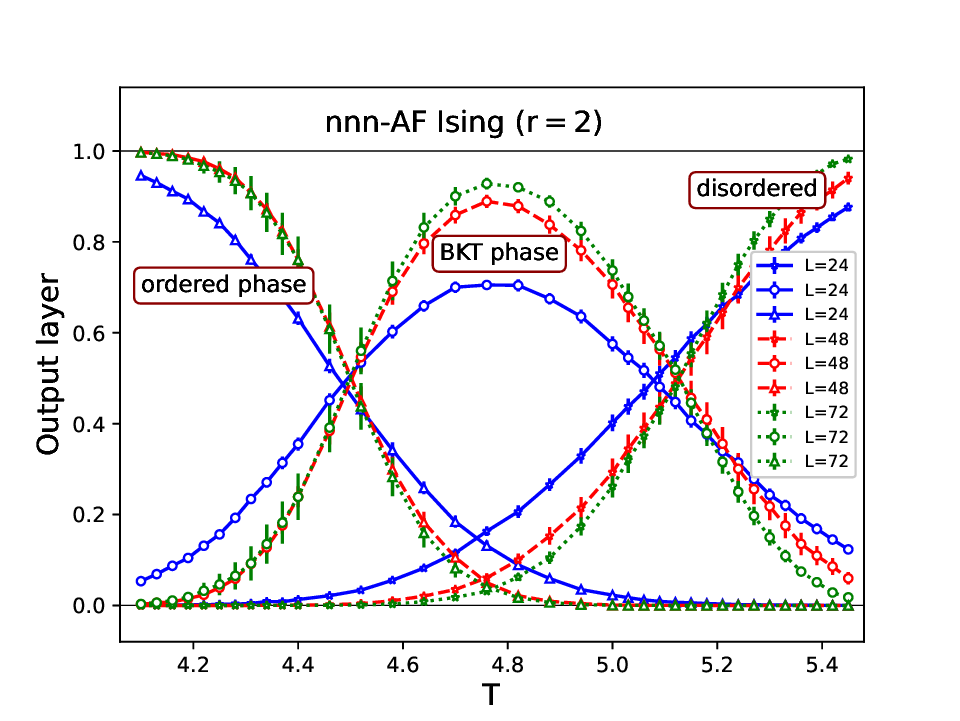}
\caption{
 The machine-learning study of the AF Ising model with F NNN coupling for $r=1/3$, 1/2, 1.0, and 2.0 on the kagome lattice. 
 The output layer, averaged over a test set as a function of $T$, is plotted using the training data of the F six-state clock model.
 The system sizes are $L = 24$, 48, and 72.
}
\label{fig_ML}
\end{center}
\end{figure}
}
\begin{document}
\title{%
 Berezinskii-Kosterlitz-Thouless phase transitions of the antiferromagnetic Ising model with ferromagnetic next-nearest-neighbor interactions on the kagome lattice
} 
\author{Yutaka Okabe}
\email{okabe@phys.se.tmu.ac.jp}
\affiliation{Department of Physics, Tokyo Metropolitan University, Hachioji, Tokyo 192-0397, Japan}
\author{Hiromi Otsuka}
\email{otsuka@tmu.ac.jp}
\affiliation{Department of Physics, Tokyo Metropolitan University, Hachioji, Tokyo 192-0397, Japan}
\date{\today}
\begin{abstract}
 We investigate the six-state clock universality of the Ising model on the kagome lattice, considering antiferromagnetic nearest-neighbor (NN) and ferromagnetic next-nearest-neighbor (NNN) interactions.
 Our comprehensive study employs three approaches: the level-spectroscopy method, Monte Carlo simulations, and a machine-learning phase classification technique. 
 In this system, we observe two Berezinskii-Kosterlitz-Thouless (BKT) transitions. 
 We present a phase diagram consisting of three phases: the low-temperature ordered phase \CHK{with sublattice magnetizations}, the intermediate BKT phase, and the high-temperature disordered phase, as a function of the ratio of the NNN interaction to the NN interaction. 
 We verify the six-state clock universality through the machine-learning study, which uses data from the six-state clock model on the kagome lattice for training.
\end{abstract}
\maketitle

\section{INTRODUCTION}
\label{sec_INTRODUCTION}
 Some effects of preventing the stabilization of ordered states have been frequently discussed in condensed matter physics. 
 Among them, the frustrations have consistently drawn considerable attention from physicists in both theoretical and experimental areas, resulting in a substantial body of research in the literature.
 
 One outstanding topic in strongly frustrated magnets is the so-called spin-ice systems, such as Ho$_2$Ti$_2$O$_7$ \cite{Harris1997} and Dy$_2$Ti$_2$O$_7$ \cite{Ramirez1999}. 
 Rare-earth ions possess a large magnetic moment and form the pyrochlore lattice, which consists of the corner-shearing tetrahedra.
 Further, because the magnetic easy axes point to the center of the tetrahedron, the spin ice is given as the antiferromagnetic (AF) Ising spin system.
 For these titanates, the long-range dipolar interactions may affect their low-temperature properties \cite{Melko2001, Isakov2005}. 
 Nevertheless, their short-range model possesses a macroscopic degeneracy in the ground state, similarly to the proton positions in water ice \cite{Pauling1935}.

 Although a degree of degeneracy in a ground state, represented by residual entropy, indicates how strongly systems are frustrated, accurate estimations are available for some restricted cases, e.g., classical spin systems defined on two-dimensional (2D) lattices.  
 Among them, the AF Ising model on the kagome lattice is of special significance in the study of frustrated magnets: it was exactly solved, and the residual entropy was calculated as $S_{\rm res}\simeq0.50183$ \cite{Kano1953}. 
 This value is larger than that of AF Ising models defined on other lattices, e.g., the triangular lattice \cite{Wannier1950} and the Villain lattice \cite{Villain1979}.
 Furthermore, the correlation length is finite~\cite{Suetoe1981} and was estimated as $\xi\simeq1.2506$ \cite{Apel2011}, so the ground state ensemble of the AF kagome Ising model is in the disordered phase.  
 This behavior is quite in contrast to that of other frustrated Ising models--- some of them possess critical ground states \cite{Stephenson1964,Villain1979}; the kagome lattice is, therefore, considered an important research ground for the effects of frustration. 

 Because temperature cannot control the entropy effects in these frustrated systems, additional interactions attract attention \cite{Chern2012,Colbois2021,Su2023}.
 For instance, Wolf and Schotte investigated the kagome AF Ising model with further neighbor interactions as a model for CsOH$\cdot$H$_2$O and found magnetically ordered phases at low temperature \cite{Wolf1988}.
 Takagi and Mekata discussed the next-nearest neighbor (NNN) interaction effects~\cite{Takagi1993}.
 In the ferromagnetic (F) NNN case, they argued that two phase transitions occur according to the six-state clock universality class, 
 \CHK{%
 and that the low-temperature ferrimagnetic phase is stabilized by frustration and F NNN coupling.}

 In this research, we revisit the kagome AF Ising model with NNN interactions. 
 Let $\Lambda$ be the kagome lattice, which consists of three interpenetrating sublattices $\Lambda_\alpha$~($\alpha=0,1,2$), i.e., $\Lambda=\Lambda_0\cup\Lambda_1\cup\Lambda_2$. 
 Then, NNN pairs of sites $\l\!\l i,j\r\!\r$ in $\Lambda$ correspond to NN pairs of sites in kagome lattices $\Lambda_\alpha$; see Fig.~\ref{fig_KagomeLattice}.
 We consider the following Hamiltonian \cite{Takagi1993}: 
 \begin{equation}
 H(J_1,J_2)
 =
 J_1\sum_{\substack{\l i,j\r \\ i,j\in\Lambda}}s_is_j
 -
 J_2\sum_{\alpha=0,1,2}\sum_{\substack{\l i,j\r \\ i,j\in\Lambda_\alpha}}s_is_j,
 \label{eq_H}
 \end{equation}
 where $s_i=\pm1$ denotes an Ising variable on the $i$th site.
 We focus on the case with AF NN interaction $J_1\ge0$ and F NNN interaction $J_2\ge0$.
 Exact results are available for the $J_2=0$ and $J_1=0$ cases \cite{Syozi1951,Kano1953}.
 These limiting models exhibit quite contrasting behaviors as temperature decreases: the former is in the disordered phase, whereas the latter exhibits a low-temperature ordered phase with an eight-fold degeneracy, possessing finite sublattice magnetizations $m_\alpha$. 
 The frustration introduced by $J_1$ to the latter lifts the degeneracy and stabilizes the six-fold ordered state with a nine-site magnetic unit-cell structure \cite{Wolf1988,Takagi1993}. 
 As a result, the ground state breaks both time-reversal and sublattice symmetries and takes the complex order parameter values as 
 \begin{equation}
 m
 =
 |m|\times e^{i2\pi(\sigma/2)}\times e^{i2\pi(\alpha/3)}
 \quad
 (\sigma=0,1).
 \label{eq_OP}
 \end{equation}
 The breaking of the emerging $\mathbb{Z}_6$ symmetry can be identified as the transition to an ordered phase in the 2D six-state clock model \cite{Wolf1988,Takagi1993,Chern2012,Su2023}, and thus is of the Berezinskii-Kosterlitz-Thouless (BKT) type \cite{Berezinskii1971,Berezinskii1972, Kosterlitz1973,Kosterlitz1974}.
 This scenario simultaneously leads to an intermediate critical phase and another BKT transition to the disordered phase at higher temperature \cite{Jose1977}.
 Some numerical simulations revealed behaviors consistent with the six-state clock universality \cite{Chern2012,Su2023}.

 In this research, we numerically investigate the Hamiltonian~(\ref{eq_H}) to elucidate its global phase diagram.
 Further-neighbor interaction effects for the triangular-lattice~\cite{Nienhuis1984,Kitatani1988,Bloete1993,Qian2004,Otsuka2006}
 and the square-lattice AF models~\cite{Grest1981,denNijs1982,Otsuka2004,Otsuka2005}
 have been studied.
 For them, we could adapt field theory, which describes the 2D Gaussian criticality in their ground-state ensembles. 
 Here, we demonstrate that an effective field theory, specifically the 2D dual sine-Gordon model, can also describe the phase transitions observed across all parameter ranges, including the off-critical ground-state ensemble of the NN AF Ising model.
 To achieve this aim, we first perform diagonalization calculations of the row-to-row transfer matrices and then analyze their eigenvalue structures based on the so-called level spectroscopy method (LS) \cite{Nomura1994,Nomura1995}.
 Although the transfer-matrix calculation imposes strong limitations on tractable system sizes, the method efficiently explores phase transitions in the entire model-parameter region \cite{Otsuka2005,Otsuka2006}.
 Second, we conduct large-scale Monte Carlo (MC) simulations to estimate the BKT transition temperatures. We also employ a newly developed machine-learning approach to investigate the six-clock universality class. 
 Their estimates of the BKT transition temperatures are compared with those obtained by the transfer-matrix calculation and with data available in the literature \cite{Su2023}.
 
 The importance of verifying six-state clock universality is that the 2D sine-Gordon theory can provide an effective description of the phase transitions in the AF kagome spin system. 
 In particular, we can successfully apply the LS method to the super-frustrated model for the first time \cite{Suetoe1981}.
 Thus, the verification implies the wider applicability of our theoretical and numerical approach to frustrated spin systems with disordered ground states; see also Sec.~\ref{sec_SUMMARY}.

 This paper is organized as follows: 
 In Sec.~\ref{sec_NUMERICAL}, we explain the details of the numerical transfer-matrix calculation method, the MC simulation method, and the machine-learning approach. 
 In Sec.~\ref{sec_RESULTS}, we provide numerical calculation results that include the global phase diagram of Eq.~(\ref{eq_H}).
 Comparisons of the BKT transition temperatures obtained by these methods are presented.
 We also refer to some transition temperatures in the literature to check the accuracy of our estimates.
 Section~\ref{sec_SUMMARY} is devoted to the summary and discussion. 

 \myfigOne

\section{NUMERICAL METHODS}
\label{sec_NUMERICAL}
\subsection{Level-spectroscopy in transfer-matrix calculations}

 Here, we summarize the methodological aspects of the numerical transfer-matrix (TM) calculations. 
 The partition function of Eq.~(\ref{eq_H}) is given by means of the row-to-row transfer matrices as 
 \begin{equation}
  Z
  =
  \Tr~e^{-H(J_1,J_2)/T}
  =
  \lim_{N\to\infty}\Tr\left({\bf T}_{\rm even}{\bf T}_{\rm odd}\right)^{\frac{N}2}.
 \end{equation}
 We suppose periodic boundary conditions in both the horizontal and vertical directions and set $\kB=1$. 
 We define the elements of the TMs, ${\bf T}_{\rm even}(s,s')$, to connect an $L$ Ising-spin configuration $s=\{s_1,\cdots,s_L\}$ on one row to $s'=\{s'_1,\cdots,s'_L\}$ on the next row; see Fig.~\ref{fig_KagomeLattice} and below.
 To this end, we perform a reduction calculation over the intermediate $L/2$ Ising spin configurations $\bar{s}=\{\bar{s}_1,\cdots,\bar{s}_{\frac{L}2}\}$, for example, as
 \begin{equation}
 {\bf T}_{\rm even}(s,s')
 =
 {\bf U}(s,s') \sum_{\bar{s}}{\bf V}_{\rm even}(s,\bar{s}){\bf V}_{\rm even}(s',\bar{s}),
 \label{eq_T}
 \end{equation}
 which requires significant numerical effort.
 Note that, as introduced in \cite{Kitatani1988}, one may be able to avoid this time-consuming process by employing an anisotropic NNN F couplings, which corresponds, in the present case, to the condition $ {\bf U}(s,s') =1$.  
 Meanwhile, as discussed in \cite{Otsuka2006}, anisotropic couplings possibly introduce one more free parameter in an effective description of phase transitions, which makes analyzing eigenvalues more complicated. 
 Therefore, we shall perform the reduction calculation as defined in Eq.~(\ref{eq_T}).
 Figure~\ref{fig_KagomeLattice} depicts a system of 5 rows $\times$ 12 columns. 
 The transfer matrix ${\bf T}={\bf T}_{\rm even}{\bf T}_{\rm odd}$ thus connects next-nearest neighboring rows and is invariant under several symmetry operations, e.g., (i) translation by two lattice spacings in the horizontal direction:~$\{s_i\}\mapsto\{s_{i+2}\}$, (ii) spin reversal symmetry:~$\{s_i\}\mapsto \{-s_i\}$, and (iii) space inversion symmetry:~$\{s_i\}\mapsto \{s_{L+1-i}\}$. 
 We perform numerical diagonalization to analyze the eigenvalue structure of ${\bf T}$.
 In this process, we utilize indices for these symmetry operations, i.e., (i) a wave number $k=2\pi q/(L/2)$, (ii) spin reversal parity ${\cal T}=\pm1$, and (iii) space inversion parity ${\cal P}=\pm1$ to characterize the excitation levels.
 
 As explained in Sec.~{\ref{sec_INTRODUCTION}}, it was clarified for some values of ratio $J_2/J_1$ that the model~(\ref{eq_H}) exhibits the two BKT transitions, say $T_1<T_2$, and belongs to the six-state clock universality class. 
 Meanwhile, our aim in this research is to clarify its global phase diagram.
 In previous publications~\cite{Otsuka2005,Otsuka2006}, we successfully applied the LS method \cite{Nomura1994,Nomura1995} for the same purpose. 
 Therefore, we also employ the LS method for this subject. 

 To make the explanations concrete, we shall borrow some notations from there.
 Based on a dual sine-Gordon model, an effective field theory for the lattice models (see Eq.~(2) in \cite{Otsuka2005} and Eq.~(4) in \cite{Otsuka2006}), we analyzed the scaling dimensions of some local operators; we denote the periodic Gaussian field and its dual as $\sqrt2\phi$ and $\sqrt2\theta$, respectively.

 Near the BKT transition at $T_1$, the so-called ${\cal M}$ operator (a part of the Gaussian term) hybridizes with the phase locking potential in the dual sine-Gordon model, $\sqrt2\cos6\sqrt2\phi$, which stabilizes the six-fold ordered states and yields the ${\cal M}$-like operator -- we shall denote its dimension as $x_0$ \cite{Nomura1995}.
 Note that the above-mentioned complex order parameter~(\ref{eq_OP}) is given in terms of the order field as $m\propto e^{i\sqrt2\phi}$.
 We also focus on a uniform magnetization operator $S:=\sqrt2\cos3\sqrt2\phi$, with scaling dimension $x_S$.
 In the numerical analysis of transfer matrices, we consider a lattice model with $L\times\infty$ cylindrical geometry.
 Then, we can estimate the renormalized scaling dimensions, e.g., $x_0(l)$ and $x_S(l)$ ($l=\ln L$), from the eigenvalues of ${\bf T}$ in the subspace specified by indices $k$, ${\cal T}$, and ${\cal P}$, denoted $\lambda_n(k, {\cal T}, {\cal P})$ \cite{Otsuka2005,Otsuka2006}.
 For instance, when we write the $n$th level as $E_n(k, {\cal T}, {\cal P})=-T\ln|\lambda_n(k, {\cal T}, {\cal P})|$, it follows that $x_0(l)\propto L[E_1(0,+1,+1)-E_0(0,+1,+1)]$ and $x_S(l)\propto L[E_0(0,-1,+1)-E_0(0,+1,+1)]$. 
 Therefore, finite-size estimates of the BKT transition temperature, $T_1(L)$, can be obtained via the level-crossing condition
 \begin{equation}
  x_0(l)=4x_S(l).
  \label{eq_LS1}
 \end{equation}

 Near the higher-temperature BKT transition $T_2$, the ${\cal M}$ operator hybridizes with a disorder operator $\sqrt2\cos\!\sqrt2\theta$ realizing the disordered phase and yields another ${\cal M}$-like operator (we denote its dimension as ${\bar x}_0$); we can find the corresponding level in the same subspace as $x_0$.
 We obtain finite-size estimates, , using the condition \cite{Otsuka2005,Otsuka2006}
 \begin{equation}
  {\bar x}_0(l)=\frac{16}9x_S(l).
  \label{eq_LS2}
 \end{equation}
 Note that the BKT transition at $T_2$ is the same as the vortex-antivortex unbinding transition observed in the usual 2D XY models \cite{Kosterlitz1973, Kosterlitz1974}.
 Vortices in the low-temperature state are strongly bound, whereas vortices are spontaneously thermally produced in the high-temperature state; see, for example, Ref.~\cite{Blundell2001}.

 Finally, we extrapolate the finite-size estimates, $T_1(L)$ and $T_2(L)$, to the thermodynamic limit according to the universal finite-size correction of $O(L^{-2})$ stemming from the descendant level of the ground state \cite{Nomura1994}; see Sec.~\ref{sec_RESULTS}. 
 In this process, logarithmic corrections from marginal operators are absent owing to the conditions employed \cite{Nomura1994,Nomura1995}, which is a crucial point in level-spectroscopy calculations.

 \subsection{Monte Carlo and machine learning calculations}

 We conduct MC simulations of the kagome AF Ising model with NNN interaction~(\ref{eq_H}). To overcome the problem of long autocorrelation times near the critical temperature. 
 We use the replica exchange method of parallel tempering~\cite{Hukushima1996}. 
 
 We calculate the spin-spin correlation function $g(r)=\l s_is_{i+r}\r$, and take the ratio of $g(r)$ at different distances; we choose $L/2$ and $L/4$ as the two distances and define the correlation ratio
 \begin{equation}
 R(T)=\frac{g(L/2)}{g(L/4)}.
 \end{equation}
 Then, as in the Binder ratio~\cite{Binder1981}, it has a single scaling variable for finite-size scaling (FSS)
 \begin{equation}
  R(T) = \tilde f(L/\xi).
 \end{equation}
 Here, $\xi$ denotes the correlation length. 
 At the critical region, where the correlation length $\xi$ is infinite, the correlation ratio does not depend on the system size $L$. 
 Thus, we expect data collapse for different sizes in the BKT phase. 
 Above the upper BKT temperature, $T_2$, and below the lower BKT temperature, $T_1$, the data for different sizes begin to separate.
 In the case of the BKT transition, the correlation length diverges as
 \begin{equation}
   \xi \propto \exp \Big( c_{1,2}/\sqrt{|T-T_{1,2}|} \ \Big).
 \label{eq_xi}
 \end{equation}

 Another approach is the machine-learning study, developed by Shiina {\it et al.}~\cite{Shiina2020}, to classify the ordered, the BKT, and the disordered phases for the clock models.
 This work extends and generalizes the research conducted by Carrasquilla and Melko~\cite{Carrasquilla2017}, who proposed a complementary paradigm to the traditional Monte Carlo approach. 
 Using large data sets of spin configurations, they classified and identified a high-temperature paramagnetic phase and a low-temperature ferromagnetic phase. 
 This process was similar to image classification, such as the detection of handwritten numbers, using machine learning. 
 They demonstrated the use of fully connected and convolutional neural networks for the study of the 2D Ising model and an Ising lattice gauge theory.
 Since the original method by Carrasquilla and Melko~\cite{Carrasquilla2017} was limited to the case of the Ising model, Shiina {\it et al.}~\cite{Shiina2020} extended the approach.
 Rather than treating the spin configuration itself, they treated a long-range spatial correlation configuration with a distance of $L/2$.
 Various spin models, including multi-component systems and systems with a vector order parameter, have been treated similarly. 
 This study examines second-order and first-order transitions, as well as the BKT transition on the same footing.

 Here, we pay special attention to verifying the six-clock universality class. 
 For the training data, we use the data from the F six-state clock model. 
 We collected typical configurations from each phase of the F clock model: ordered, BKT, and disordered. 
 We then evaluated the test data of the AF Ising model with NNN interactions across all temperatures. 
 We implemented a fully connected neural network using the standard TensorFlow library with a 100-hidden-unit model. 
 A cross-entropy cost function was employed, supplemented with an L2 regularization term to avoid the overfitting problem.
 The neural networks were trained using the Adam optimization method~\cite{Kingma2014}.
 
\section{RESULTS}
\label{sec_RESULTS}
\subsection{Level-spectroscopy calculation results}
\label{subsec_LS}

 Before presenting our results, we provide some comments on the numerical calculations.
 Compared to other numerical methods, e.g., MC simulations, exact diagonalisation calculations generally suffer from finite-size effects. 
 In our case, the low-temperature ordered phase \CHK{with sublattice magnetizations} exhibits six-site periodicity in the row direction, so the linear dimension $L$ should be a multiple of six to be commensurate with that periodicity. 
 This is why we treat transfer matrices of systems with $L=12$, 18, and 24. 
 Also, due to the NNN coupling in the vertical direction, we must trace out the $L/2$ spin degrees of freedom on an intermediate row to estimate each element of the transfer matrices; see above.
 We perform parallel computation of elements and use swap files during the diagonalization process.
 As depicted in Fig.~\ref{fig_KagomeLattice}, the transfer matrix ${\bf T}$ is not symmetric, so we use the Arnoldi iteration method. 
 Although not symmetric, the principal eigenvalues in the specified subspaces are real numbers. 

 \myfigTwo

 For clarity, we first provide the global phase diagram of the Hamiltonian~(\ref{eq_H}) in Fig.~\ref{fig_phase}. 
 Previous research has not studied such a global phase diagram based on calculations for several $r$'s.
 For this plot, we employ the Boltzmann weights $u:=e^{-J_1/T}$ and $v:=e^{+J_2/T}$ as axes of the phase diagram; the $u$ axis corresponds to the kagome-lattice AF Ising model and the $v$ axis to three decoupled kagome-lattice F Ising models. 
 To draw a phase diagram, we performed calculations for a fixed ratio, $r=J_2/J_1$, ranging from 1/8 to 8. As the temperature increased, we observed two phase transitions: one from the ordered phase to the critical (BKT) phase, and one from the critical phase to the disordered phase. 
 The case of $r=1/3$ is shown by a dotted line in Fig.~\ref{fig_phase} as an example. 
 The phase separation points are denoted by $T_1$ and $T_2$. 
 In the language of the renormalization group, the intermediate critical phase is sometimes referred to as the critical line instead of the critical point. 
 We will discuss the details of the calculation to determine $T_1$ and $T_2$ later. 
 We note a limiting case. When $J_1=0$, the system becomes a F Ising model on the kagome lattice with coupling $J_2$. 
 This model exhibits a second-order phase transition, and the critical temperature was exactly solved as $T_c/J_2 =4/\ln(3+2\sqrt{3})$ \cite{Syozi1951}. 
 In the $u-v$ plane, $u=1$ and $v_c=e^{J_2/T_c}=(3+2\sqrt{3})^{1/4}=1.595$, which is denoted by a cross mark. 
 In the limit of $r \to \infty \ (u \to 1)$, two phase-separating temperatures $T_1$ and $T_2$ merge into a single point. 
 The total $u-v$ phase diagram can be summarized as follows: The kagome-lattice AF Ising model with NNN F coupling exhibits two BKT transitions and belongs to the six-state clock universality class except at $u = 1$. 
 One may notice that the triangular-lattice AF Ising model with NNN F couplings and also the square-lattice AF three-state Potts model with NNN F couplings possess phase diagrams with a similar structure to Fig.~\ref{fig_phase}; see Fig.~3 in \cite{Otsuka2006} and Fig.~2 in \cite{Otsuka2005}.
 However, unlike those cases, the disordered phase is stabilised on the $v$ axis, which is characteristic of the kagome-lattice case.
 
 \myfigThree
 
 Here, we provide the details of calculations at $r=1/3$ as an example.    
 We plot the finite-size estimates of $T_1$ and $T_2$ and their extrapolation to the thermodynamic limit $L\to\infty$ in Fig.~\ref{fig_extrap1}.
 The insets depict how the level-crossing conditions Eqs.~(\ref{eq_LS1}) and (\ref{eq_LS2}) are satisfied by the relevant levels for $L=24$. 
 In both cases, the dimension of the uniform magnetic operator $x_{S}$ is an increasing function of $T$. 
 However, the dimension of the ${\cal M}$-like operator $x_0$ ($\bar{x}_0$) is a weakly increasing (decreasing) function around $T_1$ ($T_2$). 
 This shows the hybridization of $\sqrt2\cos6\sqrt2\phi$ ($\sqrt2\cos\sqrt2\theta$) with the ${\cal M}$ operator as expected. 

 \myfigFour

 Then, according to the $O(L^{-2})$ finite-size corrections, we extrapolate the $L=18$ and 24 data to the thermodynamic limit; see the fitting lines in Fig.~\ref{fig_extrap1}.
 Despite the small sizes, the linearity of the dependence is good, which indicates an absence of logarithmic corrections in the finite-size estimates \cite{Nomura1994, Nomura1995}.
 Comparing the results for the BKT transition temperatures to those from MC data and previous estimates \cite{Su2023}, we find they are consistent (see below).

 Next, we explore the phase transition along the $v$ axis, where each elementary triangle satisfies the one-up-two-down or two-up-one-down condition, providing the off-critical ground state ensemble; see Sec.~\ref{sec_INTRODUCTION}. 
 Assuming the same scenario as above, we search for the values of $v$ satisfying Eqs.~(\ref{eq_LS1}) and (\ref{eq_LS2}).
 Figure~\ref{fig_extrap2} plots the data in a manner similar to Fig.~\ref{fig_extrap1}. 
 We observe the level crossings (see the insets) and obtain finite-size estimates of the BKT transition points $v_1(L)$ and $v_2(L)$. 
 Figure~\ref{fig_phase} plots their extrapolated values on the $v$ axis and reveals that they join the smooth phase boundary lines estimated for finite $r$ values. 

 \myfigFive
 
 As a result, except for the decoupled case $u=1$, we observe the intermediate critical phase, which separates the disordered and ordered phases. 
 The 2D criticality stabilized in the present model~(\ref{eq_H}) is characterized by conformal field theory (CFT) with central charge $c=1$ \cite{Belavin1984}. 
 Figure~\ref{fig_c} shows the system-size dependence of the largest eigenvalue of the transfer matrix $E_0(0,+1,+1)$ for $r=1/3$ and $T=0.85$ \cite{Bloete1986, Affleck1986}; see the double circle in Fig.~\ref{fig_phase}.
 The linearity of the system-size dependence of the energy is good; from the slope of the line in Fig.~\ref{fig_c}, we estimate $c\simeq1.03$, which supports the effective description based on $c=1$ CFT. 

\subsection{Monte Carlo and machine learning calculation results}
\label{subsec_MCML}

 The MC simulations of the kagome AF Ising model with NNN interactions~(\ref{eq_H}) were performed for the coupling ratios $r=1/3$, 1/2, 1.0, and 2.0; the results are presented in Fig.~\ref{fig_MC}. 
 The correlation ratio, $R(T)$, is plotted in the left panels, where the system sizes are $L=48$, 72, 96, and 144.
 We note that the spins at distances $L/2$ and $L/4$ to calculate the correlation functions $g(L/2)$ and $g(L/4)$ belong to the same sublattice, $\Lambda_\alpha$.
 One can see that the curves for different sizes collapse at intermediate temperature regimes and spread out at lower and higher temperatures.
 The divergence of the curves at lower and higher temperatures indicates the occurrence of BKT transitions.
 
 Based on the exponential divergence of the correlation length, Eq.~(\ref{eq_xi}), we perform the finite-size-scaling (FSS) analysis as shown in the right panels of Fig.~\ref{fig_MC}, where the scaled variable is $X(c, T)=L/\exp(c_{1,2}/\sqrt{|T-T_{1,2}|})$.
 These plots provide FSS estimates of the BKT temperatures, $T_1$ and $T_2$; we summarize the results in Table~\ref{table_BKT}.

  Next, we present the findings of the machine-learning investigation of the model~(\ref{eq_H}).
  We prepared for the training data of the F six-state clock model to classify the ordered, BKT, and disordered phases. 
  Samples of spin configurations at temperature in the ranges
 $0.51\ge T_{\rm clock}\ge 0.60$,
 $0.69\ge T_{\rm clock}\ge 0.78$, and
 $0.88\ge T_{\rm clock}\ge 0.97$ are used as training data for three phases because $T_1$ and $T_2$ of the kagome-lattice F six-state clock model are estimated as 0.645 and 0.835, respectively~\cite{Okabe2025}.
 
 We then test the spin configuration of the kagome AF Ising model with NNN 
interactions. 
 The output layers averaged over a test set as a function of temperature $T$ 
are shown in Fig.~\ref{fig_ML} for the coupling ratios $r=1/3$, 1/2, 1.0, and 2.0.
 The system sizes are 24, 48, and 72.
We note again that the spins used to calculate the spatial correlation with a distance $L/2$ are in the same sublattice, $\Lambda_\alpha$.
 This figure illustrates the predicted probabilities of the phases as a function of temperature.
 Three distinct phases can be discerned: the ordered phase, the BKT phase, and the disordered phase.
 The size-dependent $T_{1,2}(L)$ is estimated from the point at which the probabilities of predicting two phases are equal to 50\%.
 We also summarize the approximate values of $T_{1,2}$ in Table~\ref{table_BKT}. 

\myfigSix

\myfigSeven

\begin{table}
\caption{
The estimates of $T_{1,2}$ for the AF Ising model with F NNN coupling on the kagome lattice
by the level-spectroscopy method (LS), the MC simulation method (MC), and the machine-learning method (ML).
}
\label{table_BKT}
\begin{center}
\begin{tabular}{lllllllll}
\hline
\hline
$r$ \
& \multicolumn{2}{l}{~~~~1/3} 
& \multicolumn{2}{l}{~~~~1/2} 
& \multicolumn{2}{l}{~~~~1.0} 
& \multicolumn{2}{l}{~~~~2.0}\\
& $T_1$ &  $T_2$ \
& $T_1$ &  $T_2$ \
& $T_1$ &  $T_2$ \
& $T_1$ &  $T_2$ \\
\hline
LS \ &  0.800 &  0.934 \ &  1.197 &  1.390 \ &  2.365 &  2.688 \ &  4.634 & 5.126 \ \\
MC \ &  0.818 &  0.936 \ &  1.214 &  1.388 \ &  2.410 &  2.705 \ &  4.778 & 5.118 \ \\
ML \ &  0.78  &  0.97  \ &  1.15  &  1.40  \ &  2.31  &  2.75  \ &  4.50  & 5.12  \ \\
\hline
\hline
\end{tabular}
\end{center}
\end{table}

\subsection{Comparison of numerical estimates of $T_{1,2}$}

 We are now able to compare the BKT transition temperatures, $T_{1,2}$, estimated in this and previous research \cite{Takagi1993,Su2023}.

 Table~\ref{table_BKT} give the estimates of $T_{1,2}$ obtained in Secs.~\ref{subsec_LS} and \ref{subsec_MCML}.
 One finds that for various values of $r$, the predictions by "MC" and "LS" are consistent, while the BKT phase of "ML" is slightly broader than those of the other two. 
 This tendency of broadening the intermediate critical phase was also observable in the machine-learning study of original six-state clock models on the square lattice~\cite{Shiina2020} and on other 2D lattices~\cite{Okabe2025}.
 This machine-learning method uses information on the entire spin configuration, so it cannot handle large sizes, and it does not sufficiently incorporate logarithmic divergence for BKT transitions.
 Therefore, it is not as effective as other sophisticated methods in determining quantitative transition temperatures.
 The advantage of this method is that it can directly verify six-state clock universality based on the similarity of configurations in completely different spin models.
 We comment on other studies. There have been other attempts to apply machine learning to the BKT transition problem \cite{Miyajima2021,Ng2023,Haldar2024}.
 Miyajima {\it et al.}~\cite{Miyajima2021} discussed two BKT transitions in a clock model. 
 They observed that $T_1$ and $T_2$ exhibited broadening of the intermediate critical phase for the eight-state clock model. 
 Although the procedure differs, the situation is similar to ours.
 
 Now, we quote the early research data obtained by Takagi and Mekata using critical exponents of spin correlation function~\cite{Takagi1993}: $T_1=1.1$ and $T_2=1.4$ for $r=1/2$.
 Our estimates are $T_1=1.197$ and $T_2=1.390$ using the LS method. 
 They had predicted transition temperatures compatible with our results. 
 This work pioneers the proposal of a six-state clock universality. 
 However, the size of the simulation they performed was rather small ($L \le 66, N=(4/3)L^2$), and they did not employ sophisticated analysis, such as FSS. Therefore, the numerical accuracy is not very good.
 Also we quote the recent data obtained by Su {\it et al.} using the distributions of two-configuration Hamming distances; see Table.~I in Ref.~\cite{Su2023}:
 $T_1=0.806$ and $T_2=0.930$ for $r=1/3$. 
 They performed simulations up to $L=120, N=3L^2$, and they used a FSS analysis. This size is of the same order as our MC size ($L \le 144, N=(3/2)L^2$). 
 Therefore, the numerical accuracy is the same as ours.
 We confirm their results are quite close to our data in Table~\ref{table_BKT}.
 These consistency checks confirm the accuracy of the global phase diagram presented in Fig.~\ref{fig_phase}. 

\section{SUMMARY AND Discussion}
\label{sec_SUMMARY}
 We have investigated the NN AF Ising model with NNN F coupling defined on the kagome lattice.
 Although the NN case is in the disordered phase, the Boltzmann weight from the NNN coupling tends to decrease the model's effective temperature, stabilizing first the intermediate critical phase and then the six-fold ordered phase. 
 Using three approaches --- LS, MC, and ML --- we presented the global phase diagram, consisting of three phases: the low-temperature ordered phase \CHK{with sublattice magnetizations}, the intermediate BKT phase, and the high-temperature disordered phase. 
 We verified six-state clock universality through a machine-learning study that trained on data from the six-state clock model on the kagome lattice.
 
 The AF kagome Ising model is sometimes called ``super-frustrated'' \cite{Suetoe1981} to emphasize its disordered ground state.
 Meanwhile, ``normally frustrated'' models, for example, the NN AF triangular Ising model, possess a critical ground state ensemble \cite{Stephenson1964}, which is mapped to a solid-on-solid model \cite{Nienhuis1984}.
 In such a case, a continuum effective theory is a 2D Gaussian model, so a sine-Gordon model can be used to analyze the instabilities of the critical ensemble \cite{Bloete1982,Bloete1993}.
 Returning to the kagome case, we have observed that the disordered ensemble can be described by the 2D Gaussian model perturbed by the relevant defect field $\sqrt2\cos\sqrt2\theta$, and thus the dual sine-Gordon model also explains how the present model exhibits phase transitions.

 It is interesting to discuss how a microscopic description of vortices can be applied to the current AF Ising model with NNN F coupling on the kagome lattice. 
 One approach to exploring this issue is to investigate the excited state structure from the ground state using, for instance, the Wang-Landau MC method \cite{Wang2001}. 
 This topic will be addressed in future work.
 
 Now, we comment on the relevance to spin ice systems \CHK{\cite{Harris1997,Ramirez1999,Bramwell2001}}. 
 The 2D analogue of spin ice, kagome ice, was found in the pyrochlore compound Dy$_2$Ti$_2$O$_7$ in a magnetic field along the [111] direction \cite{Matsuhira2002};
 \CHK{see also~\cite{Wills2002}}.
 Kagome ice is the plateau state with 1/3 of the saturation magnetization and corresponds to an NN AF Ising model on the kagome lattice.
 The ground state of kagome ice possesses macroscopic degeneracy, whose spin configurations can be mapped to dimer covering patterns of the honeycomb lattice \cite{Udagawa2002,Moessner2003}. 
 Thus, it is in the critical phase~\cite{Moessner2003}, and preserves sublattice symmetry, but exhibits uniform magnetization. 
 As mentioned, the present model cannot reproduce these properties, and this discrepancy may imply the importance of studying the magnetic field effects \cite{Wills2002}.

 As mentioned, we directly verified the six-state clock universality by applying the machine-learning approach developed by Shiina {\it et al.}~\cite{Shiina2020}. 
 The present model of a frustrated AF system with NNN interaction exhibits a two-step transition, which suggested six-clock universality class.  Therefore, we selected a simple F clock model as the training data and attempted the phase classification.
 This method is widely applicable to other models.
 It is also an interesting problem to select proper training data without using prior knowledge. 

\begin{acknowledgments}
 The authors thank Kenta Shiina, Hiroyuki Mori, and Hwee Kuan Lee for valuable discussions.
 In numerical diagonalization calculations using the Arnoldi method, we employed Spectra, a C++ library available at https://github.com/yixuan/spectra/.
 JSPS KAKENHI Grant Number JP22K03472 supported this work. 
\end{acknowledgments}
%

\end{document}